# Counting statistics based on the analytic solutions of the differential-difference equation for birth-death processes

Seong Jun Park[a,b] and M.Y. Choi[b]

[a] National CRI-Center for Chemical Dynamics in Living Cells, Chung-Ang University, Seoul 06974, Korea

[b] Department of Physics and Astronomy and Center for Theoretical Physics, Seoul National University, Seoul 08826, Korea

## Abstract

Birth-death processes take place ubiquitously throughout the universe. In general, birth and death rates depend on the system size (corresponding to the number of products or customers undergoing the birth-death process) and thus vary every time birth or death occurs, which makes fluctuations in the rates inevitable. The differential-difference equation governing the time evolution of such a birth-death process is well established, but it resists solving for a non-asymptotic solution. In this work, we present the analytic solution of the differential-difference equation for birth-death processes without approximation. The time-dependent solution we obtain leads to an analytical expression for counting statistics of products (or customers). We further examine the relationship between the system size fluctuations and the birth and death rates, and find that statistical properties (variance subtracted by mean) of the system size are determined by the mean death rate as well as the covariance of the system size and the net growth rate (i.e., the birth rate minus the death rate). This work suggests a promising new direction for quantitative investigations into birth-death processes.

Email addresses: tcpsj123@cau.ac.kr (Seong Jun Park); mychoi@snu.ac.kr (M.Y. Choi)

## 1. Introduction

A variety of quantities in natural and social phenomena undergo a birth-death process. The theory of birth-death processes was developed at the beginning of the twentieth century to analyze population growth. Over time, it has become complicated, spawning new branches of stochastic processes [1-4], including biomolecules, such as DNA or protein, bacteria, people with a disease, or customers in line at a supermarket [5-7].

A birth-death process is a specific type of continuous-time Markov chain describing a stochastic process. It consists of a set of states $\{0,1,2,\cdots\}$ for which transitions from state *n* may only reach either state *n*−1 or state *n*+1, typically denoting the *population* or *size* of a system. In the birth-death process, the state represents the number of products or customers in the system. Namely, "birth" corresponds to product creation or customer arrival, and "death" to product annihilation or customer departure. In most cases, two fundamental characteristics of the birth-death process provide an adequate description of the birth-death system: (1) system capacity and (2) the number of death channels.

In a birth-death system, there may be either no limit to the number allowed at any time or a physical limitation in the system size. System capacity refers to the maximum system size. For instance, a waiting room that can accommodate ten persons represents a system of finite system capacity. When ten people are in the waiting room, no further customers can enter the room until it becomes available according to service completion. As for the number of death channels, we refer to such systems as hair-styling salons, supermarkets, or restaurants, which have finite numbers of death channels. According to the queuing theory, the primary characteristics of a birth-death process include queue discipline and the number of service stages. In this work, we limit the discussion to the effects of the system capacity and the number of death channels.

Analyzing a birth-death system often aims to determine the number of products or customers in the system at a specific moment. For example, one might be interested in the number of customers waiting in front of a checkout in a supermarket, that of infected people during a pandemic, or the count of a specific protein controlling metabolism. Accordingly, there have been a considerable number of studies on counting statistics of products (or customers). In renewal theory the time required for birth or death is considered an independently and identically distributed random variable [8, 9]. Some studies of counting statistics have focused on the effects of the batch size of the production (arrival) or destruction (departure) [10-12] on the population dynamics, extending simple renewal theory. However, in many practical applications, birth and death rate coefficients depend on the system size to which renewal theory is no longer applicable; the concern regarding system-size-dependent birth and death rates has thus been growing [13-18]. The Chemical Fluctuation Theorem (CFT), proposed recently for gene expression [19], applies to a general birth-death process in which birth is described by an arbitrary stochastic process and death by a renewal process.

Such a birth-death process is described by a differential-difference equation known as the master equation. Solving the equation provides information regarding how products or customers change over time. For some specific birth-death processes, the solution of the equation is known [7, 20-23]. In many real situations, the birth or death rate depends on the number of products or customers in the birth-death process [15, 17, 24-26]. However, the time-dependent solution to the differential-difference equation for birth-death processes remains unknown when the birth or death rate depends on the system size.

In this work, we determine the solution of the differential-difference equation for birth-death processes with the birth or death rate given by an arbitrary function of the system size. The solution helps derive the analytic formula of the product number counting statistics. (Henceforth, the term "product" may also stand for "customer", etc.; likewise, "production" and "destruction" also for "arrival" and "departure", respectively.) We also examine the relationship between product number fluctuations and birth/death rates. These theoretical results for the time evolution of the system undergoing the birth-death process, confirmed by some computations, would help understand quantitatively related natural and social phenomena and find treatments for diseases, develop drugs, establish new policies, and so on.

This paper is organized as follows: Section 2 introduces the well-known differential-difference equation for the birth-death process. Its solution is obtained and classified according to the system capacity and number of death channels. We then present the counting statistics of the product number. Also given are the analytic results, which reveal the relationship between the product number counting statistics and the birth/death rates. Finally, Section 3 presents the summary and discussion.

## 2. Birth and Death Rate Fluctuations

2.1 *Solution of the differential-difference equation for birth-death processes*

We consider a general birth-death system where the birth and death rates vary with the number of products. To describe accurately such a realistic birth-death system, we first introduce the differential-difference equation for the time evolution of the system:

$$\partial P_n(t)/\partial t = -(\lambda_n + \mu_n)P_n(t) + \lambda_{n-1}P_{n-1}(t) + \mu_{n+1}P_{n+1}(t) \quad (n \geq 0), \tag{1}$$

where $P_n(t)$ denotes the probability that there are $n$ products/customers at time $t$, $\lambda_n \, (> 0)$ is the birth rate when the system size increases from $n$ to $n+1$, and $\mu_n \, (> 0)$ is the death rate when the system size decreases from $n$ to $n-1$. The first term on the right-hand side of Eq. (1) describes the loss in $P_n(t)$ due to production/arrival increasing the product number from $n$ to $n+1$ and due to destruction/departure reducing the product/customer number from $n$ to $n-1$. The second and third terms describe the gain due to production/arrival increasing the number from $n-1$ and destruction/departure reducing the number from $n+1$. When $n = 0$, there is no product to destroy (or customer to depart), leading to $\mu_0 = 0$; on the other hand, $\lambda_0$ in general does not vanish, representing the birth rate of the first product (or the arrival rate of the first customer). There may not be a negative number of products, namely, $P_n(t) = 0$ for $n < 0$, and the normalization condition reads $\sum_{n=0}^{N} P_n(t) = 1$, where $N$ is the system capacity given by a positive integer. Equation (1) can be derived from the theory of continuous-time Markov chains or by using the concept of the flow of balance [7, 9, 27].

Solving Eq. (1) for given birth/death rates $\lambda_n$ and $\mu_n$, one can obtain the probability $P_n(t)$ provided that the initial condition $P_n(0)$ is known. In particular, $P_n(t)$ follows a Poisson distribution in the case of the infinite system capacity ($N \to \infty$) and constant birth/death rates ($\lambda_n = \lambda$ and $\mu_n = n\mu$). Note that the death process occurring at a constant rate $\mu$ gives the combined death rate $n\mu$ for the system in the presence of $n$ products. While the solution to

Eq. (1) is known for simple birth-death processes with constant birth and death rates [7, 9, 28], to our knowledge, there is no analytic expression of $P_n(t)$ available for $\lambda_n$ and $\mu_n$ given by arbitrary functions of the product number. Remarkably, even though the relationship between $P_n(t)$ and rates $\lambda_n$ and $\mu_n$ (depending explicitly on the number *n*) is not yet understood, $P_n(t)$ can be expressed in terms of only $\lambda_n$ and $\mu_n$. In short, the analytic solution of Eq. (1) is very challenging to obtain. Here we circumvent the difficulty by introducing the rate function $f_{n,m}(t)$ to complete the *m*th birth in the system of size *n*. When the system capacity is infinite ($N \to \infty$), the probability of having *n* products at time *t* is written in the form

$$P_n(t) = \frac{1}{\lambda_n} \sum_{m=n+1}^{\infty} f_{n+1,m}(t). \tag{2}$$

In the case of finite system capacity *N*, we obtain

$$\begin{cases} P_n(t) = \dfrac{1}{\lambda_n} \sum_{m=n+1}^{\infty} f_{n+1,m}(t) & (n = 0,1,2,\cdots,N-1) \\ P_N(t) = \sum_{m=0}^{\infty} \int_0^t d\tau\, f_{N,N+m}(\tau) e^{-\mu_N(t-\tau)} \end{cases}. \tag{3}$$

The Supplementary Information presents the derivations of Eqs. (2) and (3). In this work, there are four kinds of $f_{n,m}(t)$ classified by the system capacity and the number of death channels. Equations (2) and (3) satisfy the normalization conditions $\sum_{n=0}^{\infty} P_n(t) = 1$ and $\sum_{n=0}^{N} P_n(t) = 1$, respectively. Moreover, they are consistent with the previously known results for the pure birth process ($\mu_n = 0$) [9, 29] and the birth-death process with constant rates ($\lambda_n = \lambda$ and $\mu_n = n\mu$) [7, 9, 28]. Depending on the number of death channels and the system capacity, we have the solution in the form of Eqs. (2) and (3) with distinct analytic forms of $f_{n,m}(t)$. The Supplementary Information presents the complete derivation of $f_{n,m}(t)$ for each case classified according to the system capacity and the number of destruction/service channels.

2.2 *Counting statistics*

With the probability $P_n(t)$ that the system size is $n$ at time $t$, we can compute the statistics of the system size. Among the $m$th moment of a physical quantity $X$ defined to be $\langle X^m(t) \rangle \equiv \sum_{n=0}^{\infty} X^m P_n(t)$, two metrics of immediate interest are the first and second moments of the system size, which are just the mean number $\langle n \rangle$ and variance $\sigma_n^2 \equiv \langle n^2 \rangle - \langle n \rangle^2$ of products (customers); these can be derived from Eqs. (2) and (3). Multiplying both equations by $n$ or $n^2$ and summing over $n$ from 0 to $N$, which may be finite [for Eq. (2)] or infinite [for Eq. (3)], we obtain:

$$\langle n(t) \rangle = \begin{cases} \sum_{m=2}^{\infty} \sum_{k=1}^{m-1} \frac{m-k}{\lambda_{m-k}} f_{m-(k-1),m}(t) & \text{infinite system capacity } (N \to \infty) \\ \sum_{m=1}^{\infty} \sum_{k=1}^{N-1} \frac{k}{\lambda_k} f_{k+1,m+k}(t) + N P_N(t) & \text{finite system capacity } (N < \infty) \end{cases} \quad (4)$$

$$\sigma_n^2(t) = \begin{cases} \langle n(t) \rangle (1 - \langle n(t) \rangle) + \sum_{m=2}^{\infty} \sum_{k=1}^{m-1} \frac{(m-k)(m-k-1)}{\lambda_{m-k}} f_{m-(k-1),m}(t) & \text{for } N \to \infty \\ \langle n(t) \rangle (1 - \langle n(t) \rangle) + \sum_{m=1}^{\infty} \sum_{k=1}^{N-1} \frac{k(k-1)}{\lambda_k} f_{k+1,k+m}(t) + N(N-1) P_N(t) & \text{for } N < \infty \end{cases} \quad (5)$$

The most crucial point of Eqs. (4) and (5) is that one can compute the mean and variance of the system size even for the birth and death rates given by any arbitrary functions of the system size $n$. On the other hand, however, Eqs. (4) and (5) do not provide a clear relationship between the statistics of the birth and death rates ($\lambda_n$ and $\mu_n$) and the statistics of the system size.

To probe the effects of the birth and death rates on the statistics of the system size, we consider the mean system size $\langle n(t) \rangle$ expressed in terms of the birth and death rates. Multiplying both sides of Eq. (1) by $n$ and summing over $n$ results in the following equation:

$$\langle n(t) \rangle = \int_0^t [\langle \lambda_n(\tau) \rangle - \langle \mu_n(\tau) \rangle] d\tau, \quad (6)$$

which states that the derivative of the mean system size with respect to time equals the mean (net) growth rate, i.e., the difference between the mean birth rate and the mean death rate. We express the second moment of the system size in a similar manner and write Mandel's Q-parameter [30] in the form:

$$Q_n(t) \equiv \sigma_n^2(t)/\langle n(t)\rangle - 1 = 2\int_0^t \left[\rho(\tau)\sigma_n(\tau)\sigma_{\lambda_n-\mu_n}(\tau) + \langle \mu_n(\tau)\rangle\right]d\tau \Big/ \langle n(t)\rangle, \tag{7}$$

where $\sigma_n(t)$ and $\sigma_{\lambda_n-\mu_n}(t)$ are the standard deviations of the system size $n$ and the net growth rate or the birth rate minus the death rate $\lambda_n - \mu_n$, respectively. The correlation coefficient $\rho(t)$ of $n$ and $\lambda_n - \mu_n$ is defined to be the covariance of the two divided by the product of their standard deviations: $\rho(t) \equiv \text{Cov}[n, \lambda_n(t) - \mu_n(t)]/(\sigma_n(t)\sigma_{\lambda_n-\mu_n}(t))$ with the covariance given by $\text{Cov}[n, X] \equiv \langle nX\rangle - \langle n\rangle\langle X\rangle$. The statistical properties of the system size can be classified into three main groups, depending on the sign of the Q-parameter: super-Poisson $(Q_n(t) > 0)$, Poisson $(Q_n(t) = 0)$, and sub-Poisson $(Q_n(t) < 0)$. Equation (7) shows that a) the covariance between the system size $n$ and the net growth rate $\lambda_n - \mu_n$ and b) the mean death rate $\langle \mu_n(\tau)\rangle$ determine the properties of the system size. In particular, when the birth and death rates have no dependence on the system size, namely, $\lambda_n = \lambda(t)$ and $\mu_n(t) = \mu(t)$, the correlation coefficient vanishes: $\rho(t) = 0$. This in turn leads to $Q_n(t) = 2\int_0^t \langle \mu_n(\tau)\rangle d\tau \Big/ \langle n(t)\rangle > 0$, implying that the counting statistics of the system size display super-Poisson characteristics. Similarly, when the net growth rate correlates positively with the system size or an increasing function of $n$, the counting statistics of the system size possess super-Poisson characteristics: $Q_n(t) > 0$ [see Fig. 1(a)]. In contrast, when the net growth rate is inversely correlated with the system size, namely when it is a decreasing function of the system size, we can observe super-Poisson, Poisson, or sub-Poisson characteristics, depending on whether the absolute value of the covariance of the system size $n$ and the net growth rate $\lambda_n - \mu_n$ is larger than the mean death rate $\langle \mu_n(\tau)\rangle$. If the integral from zero to $t$ of the sum of the covariance and the mean death rate is greater/less than zero, the system size displays super-Poisson/sub-Poisson characteristics at time $t$ [see Fig. 1(b) and (c)]. If the integral vanishes, the system size displays Poisson characteristics. Equations (4) to (7) are consistent with the previous results for cases where only birth exists or where the birth and death rates are constant. Equations (6) and (7) remain valid regardless of the system capacity or the number of death channels.

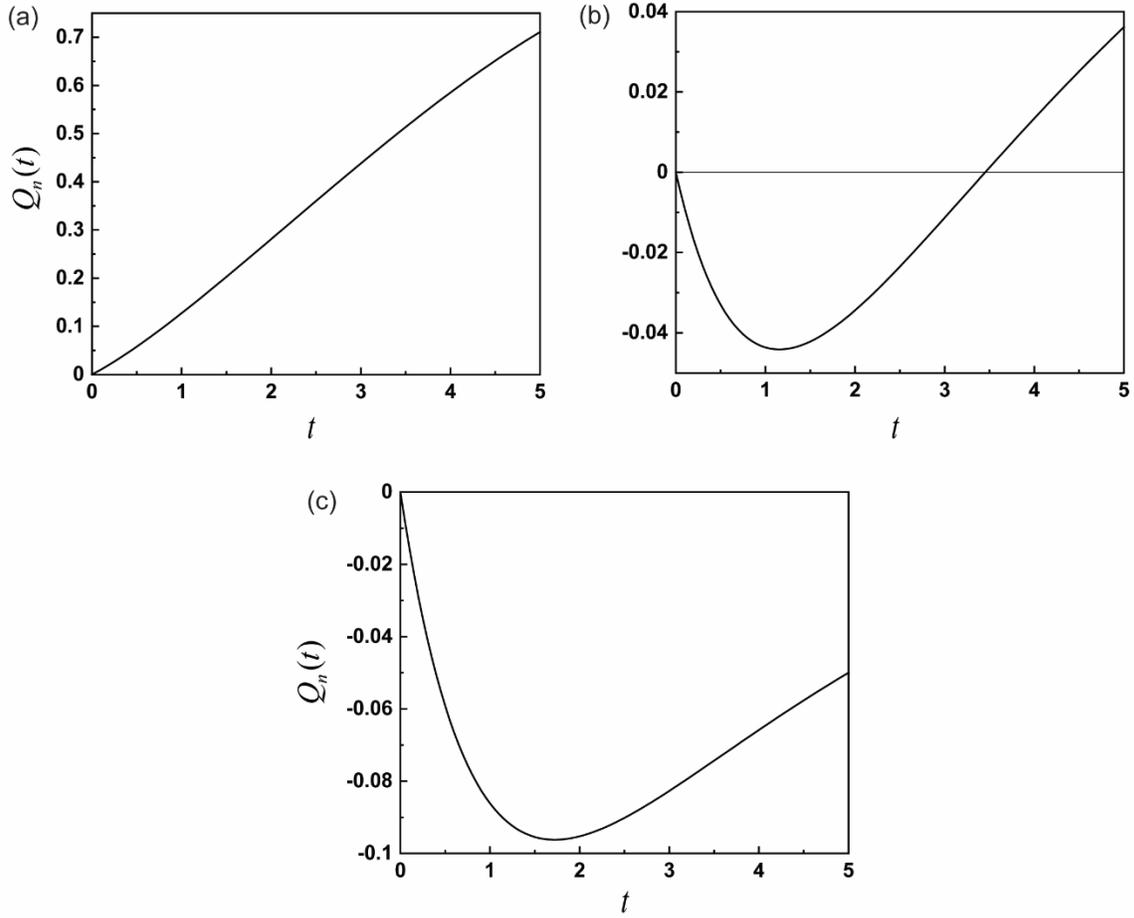

**Fig. 1.** Product number fluctuations in the birth-death system with infinite system capacity and an unlimited number of death channels. The time evolution of Mandel's Q-parameter of the product number $n$ is shown for several types of the birth rate $\lambda_n$ and death rate $\mu_n$, with different dependences on the number of product molecules. (a) $\lambda_n = 0.4 + 0.3(1-e^{-0.4n})$ and $\mu_n = 0.4 - 0.3(1-e^{-0.4n})$, with the net growth rate $\lambda_n - \mu_n$ increasing with $n$; (b) $\lambda_n = 0.4 - 0.3(1-e^{-0.4n})$ and $\mu_n = 0.4 + 0.3(1-e^{-0.4n})$, with $\lambda_n - \mu_n$ decreasing with $n$; (c) $\lambda_n = 0.4 - 0.3(1-e^{-0.8n})$ and $\mu_n = 0.4 + 0.3(1-e^{-0.8n})$, with $\lambda_n - \mu_n$ reducing with $n$. In the time domains given, the Q-parameter displays super-Poissonian characteristics $(Q_n(t) > 0)$ in (a) and sub-Poissonian characteristics $(Q_n(t) < 0)$ in (c). In (b), it exhibits both super-Poissonian and sub-Poissonian characteristics.

## 3. Summary and Discussion

We have presented how to compute the counting statistics of products undergoing a birth-death process in which birth and death rates depend on the number of products. For this purpose, we have considered the well-known differential-difference equation in Eq. (1), which mirrors the birth and death rate fluctuations with the size of the system undergoing a birth-death process. The time-dependent (non-asymptotic) solution of Eq. (1), which was hardly available in existing literature, has been determined: Specifically, the probability $P_n(t)$ that the system size is $n$ at time $t$ is given by Eqs. (2) and (3). The newly introduced rate function $f_{k,m}(t)$, describing the time required for $m$ births in the system of size $k$ (namely, when there are $k$ products), is associated with $P_n(t)$ and the product number counting statistics given by Eqs. (4) and (5).

We have also determined the relationship between product number fluctuations and birth and death rates. A net growth rate (given by the birth rate minus the death rate) increasing with the product number gives rise to super-Poisson characteristics in the product number counting statistics. In case that the net growth rate does not increase with the product number, the statistical properties at time $t$ are determined by the time integral from zero to $t$ of the sum of the mean death rate and the covariance of the system size and the net growth rate. A positive/negative value of the integral corresponds to super-/sub-Poissonian characteristics of the product number fluctuations. The statistical properties of the product number fluctuations obtained in terms of only the birth and death rates, as described by Eqs. (4) - (7), as well as the solution of the differential-difference equation for birth-death processes given by Eqs. (2) and (3) are significant achievements and would expand our understanding of birth-death processes.

Birth-death processes are observed widely in nature and society; thus, the results of this work should apply to various systems. For instance, we may consider applications to the number of infected people, the human population in specific areas, the population of fish or algae in a lake, the number of proteins that carry out a given function in a cell, and so on. Equation (2), giving the solution of Eq. (1), helps clarify the dynamics of such a birth-death system over time. This framework for extracting statistical properties of the product number provides a mathematical means for quantitative analysis of birth-death processes.

Finally, we point out that the methodology in this work can be extended to more general birth-death processes, taking into consideration, e.g., queue discipline, multiple service stages, complexity involving bulk production, etc. Those generalized birth-death processes are relevant to many fields that study population systems, such as queuing models, chemical reactions, and disease spreading [7, 28, 31-34]. The results obtained in this work are expected to serve as valuable tools for investigating them; these are left for future study.

## Acknowledgment

The authors are grateful to Prof. Jaeyoung Sung and Dr. Sangwook Lee for critical reading of the manuscript. This work was supported by the National Research Foundation (NRF) grants funded by the Korean government (Ministry of Science and ICT) [Grant No. 2021R1C1C2010450 (SJP) and No. 2022R1A2C1012532 (MYC)].

## Author Declarations

### Conflict of interest

The authors have no conflicts of interest to declare.

### Data Availability

No data were used for the research described in the article.

## Supplementary Material

**Derivation of Eqs. (2) and (3)**

We examine a system undergoing a birth-death process where birth and death rates depend on the system size, i.e., the number of products. Specifically, $\lambda_n$ denotes the birth rate when the system size increases from $n$ to $n+1$, and $\mu_n$ is the death rate when the system size decreases from $n$ to $n-1$. Based on these assumptions, we derive the probability $P_n(t)$ for the system size to be $n$ at time $t$. Before starting the derivation, we let $f_{k,m}(t)$ denote the probability rate function describing the distribution of times to complete the $m$th birth ($m = 1, 2, 3, \cdots$) in the system with size $k$ (= 1, 2, 3, ..., $m$). Note that there is no probability density unless both $k$ and $m$ are positive, namely, $f_{k,m}(t) = 0$ for $k \leq 0$ or $m \leq 0$. We now consider several cases according to the system capacity and the number of death channels.

1. *Infinite system capacity and an unlimited number of death channels*

In the birth-death system with infinite system capacity and an unlimited number of death channels, under the initial condition $P_n(0) = \delta_{n0}$, the probability rate function for the time to complete the first birth is given by $\lambda_0 e^{-\lambda_0 t}$, namely, $f_{1,1}(t) = \lambda_0 e^{-\lambda_0 t}$. When the birth is given twice (i.e., $m = 2$), the system size $k$, being not greater than $m$, takes the value one or two immediately upon the second birth, which makes it necessary to determine $f_{1,2}(t)$ and $f_{2,2}(t)$. For the population size to be one ($k = 1$) at time $t$ immediately upon the second birth, the following process should be gone through: The first birth occurs at time $\tau_1$, and the first product thus produced is destroyed during time interval $\tau_2$ without any production, and then the second birth occurs during the time interval $t - \tau_1 - \tau_2 (\geq 0)$. In accord, we can express $f_{1,2}(t)$ as

$f_{1,2}(t) = \int_0^t d\tau_1 \, f_{1,1}(\tau_1) \int_0^{t-\tau_1} d\tau_2 \, \mu_1 e^{-\mu_1 \tau_2} e^{-\lambda_1 \tau_2} \lambda_0 e^{-\lambda_0 (t-\tau_1-\tau_2)}$, which yields the Laplace transform

(LT): $\hat{f}_{1,2}(s) = \hat{f}_{1,1}(s) \dfrac{\mu_1}{s+\mu_1+\lambda_1} \dfrac{\lambda_0}{s+\lambda_0}$ with $\hat{f}(s) \equiv \int_0^\infty dt \, e^{-st} f(t)$. Next, suppose that the system size is two at time $t$, immediately upon the second birth. This corresponds to the case that the first birth occurs at time $\tau_1$, and the first product produced survives until the second birth occurs during the time interval $t-\tau_1$. We thus obtain $f_{2,2}(t) = \int_0^t f_{1,1}(\tau_1) e^{-\mu_1(t-\tau_1)} \lambda_1 e^{-\lambda_1(t-\tau_1)} d\tau_1$, together with its LT given by $\hat{f}_{2,2}(s) = \hat{f}_{1,1}(s) \dfrac{\lambda_1}{s+\mu_1+\lambda_1}$.

We then consider the case that the third production occurs. The system size must be one, two, or three immediately upon the third production. First, let us focus on the case described by $f_{1,3}(t)$, that the system size is one immediately upon the third birth. There are two possibilities for giving contributions to $f_{1,3}(t)$: One is that the system size is one immediately upon the second birth at time $\tau_1$, the only product destructs during time interval $\tau_2$, and the third birth occurs during $t-\tau_1-\tau_2$. The other possibility is that the system size is two immediately upon the second birth at time $\tau_1$, the two products destruct one by one during time intervals $\tau_2$ and $\tau_3$, and the third birth occurs during $t-\tau_1-\tau_2-\tau_3$. This results in

$$f_{1,3}(t) = \int_0^t d\tau_1 f_{1,2}(\tau_1) \int_0^{t-\tau_1} d\tau_2 \mu_1 e^{-\mu_1 \tau_2} e^{-\lambda_1 \tau_2} \lambda_0 e^{-\lambda_0(t-\tau_1-\tau_2)}$$
$$+ \int_0^t d\tau_1 f_{2,2}(\tau_1) \int_0^{t-\tau_1} d\tau_2 \mu_2 e^{-\mu_2 \tau_2} e^{-\lambda_2 \tau_2} \int_0^{t-\tau_1-\tau_2} d\tau_3 \mu_1 e^{-\mu_1 \tau_3} e^{-\lambda_1 \tau_3} \lambda_0 e^{-\lambda_0(t-\tau_1-\tau_2-\tau_3)}$$

along with its LT given by $\hat{f}_{1,3}(s) = \hat{f}_{1,2}(s) \dfrac{\mu_1}{s+\mu_1+\lambda_1} \dfrac{\lambda_0}{s+\lambda_0} + \hat{f}_{2,2}(s) \dfrac{\mu_2}{s+\mu_2+\lambda_2} \dfrac{\mu_1}{s+\mu_1+\lambda_1} \dfrac{\lambda_0}{s+\lambda_0}$.

Second, we examine $f_{2,3}(t)$, corresponding to that the system size is two immediately upon the third birth, for which there are two possibilities: One is that the system size is one immediately upon the second birth at time $\tau_1$ and the product survives during the third birth time interval $t-\tau_1$. The other is that the system size is two immediately upon the second birth at time $\tau_1$. Then, one of the two products destructs with the rate $\mu_2$ without additional production during $\tau_2$, and another survives during the third birth time interval $t-\tau_1-\tau_2$. It thus reads

$$f_{2,3}(t) = \int_0^t d\tau_1 f_{1,2}(\tau_1) \lambda_1 e^{-\lambda_1(t-\tau_1)} e^{-\mu_1(t-\tau_1)} + \int_0^t d\tau_1 f_{2,2}(\tau_1) \int_0^{t-\tau_1} d\tau_2 \mu_2 e^{-\mu_2 \tau_2} e^{-\lambda_2 \tau_2} \lambda_1 e^{-\lambda_1(t-\tau_1-\tau_2)} e^{-\mu_1(t-\tau_1-\tau_2)}$$

, and its LT is $\hat{f}_{2,3}(s) = \hat{f}_{1,2}(s) \dfrac{\lambda_1}{s+\mu_1+\lambda_1} + \hat{f}_{2,2}(s) \dfrac{\mu_2}{s+\mu_2+\lambda_2} \dfrac{\lambda_1}{s+\mu_1+\lambda_1}$. Finally, we evaluate the case that the system size is three immediately upon the third birth. There is only one possibility for this: The system size is two immediately upon the second birth at time $\tau_1$, and the two products survive during the third birth time interval $t-\tau_1$. In accord, we have

$$f_{3,3}(t) = \int_0^t d\tau_1 f_{2,2}(\tau_1) \lambda_2 e^{-\lambda_2(t-\tau_1)} e^{-\mu_2(t-\tau_1)}$$ as well as its LT in the form

$$\hat{f}_{3,3}(s) = \hat{f}_{2,2}(s) \frac{\lambda_2}{s+\mu_2+\lambda_2}.$$

In a similar fashion, one can derive $f_{k,4}(t)$ for $k = 1, 2, 3,$ and 4:

$$f_{1,4}(t) = \int_0^t d\tau_1 f_{1,3}(\tau_1) \int_0^{t-\tau_1} d\tau_2 \mu_1 e^{-\mu_1\tau_2} e^{-\lambda_1\tau_2} \lambda_0 e^{-\lambda_0(t-\tau_1-\tau_2)}$$

$$+ \int_0^t d\tau_1 f_{2,3}(\tau_1) \int_0^{t-\tau_1} d\tau_2 \mu_2 e^{-\mu_2\tau_2} e^{-\lambda_2\tau_2} \int_0^{t-\tau_1-\tau_2} d\tau_3 \mu_1 e^{-\mu_1\tau_3} e^{-\lambda_1\tau_3} \lambda_0 e^{-(t-\tau_1-\tau_2-\tau_3)}$$

$$+ \int_0^t d\tau_1 f_{3,3}(\tau_1) \int_0^{t-\tau_1} d\tau_2 \mu_3 e^{-\mu_3\tau_2} e^{-\lambda_3\tau_2} \int_0^{t-\tau_1-\tau_2} d\tau_3 \mu_2 e^{-\mu_2\tau_3} e^{-\lambda_2\tau_3} \int_0^{t-\tau_1-\tau_2-\tau_3} d\tau_4 \mu_1 e^{-\mu_1\tau_4} e^{-\lambda_1\tau_4} \lambda_0 e^{-\lambda_0(t-\tau_1-\tau_2-\tau_3-\tau_4)}$$

$$f_{2,4}(t) = \int_0^t d\tau_1 f_{1,3}(\tau_1) \lambda_1 e^{-\lambda_1(t-\tau_1)} e^{-\mu_1(t-\tau_1)}$$

$$+ \int_0^t d\tau_1 f_{2,3}(\tau_1) \int_0^{t-\tau_1} d\tau_2 \mu_2 e^{-\mu_2\tau_2} e^{-\lambda_2\tau_2} \lambda_1 e^{-\lambda_1(t-\tau_1-\tau_2)} e^{-\mu_1(t-\tau_1-\tau_2)}$$

$$+ \int_0^t d\tau_1 f_{3,3}(\tau_1) \int_0^{t-\tau_1} d\tau_2 \mu_3 e^{-\mu_3\tau_2} e^{-\lambda_3\tau_2} \int_0^{t-\tau_1-\tau_2} d\tau_3 \mu_2 e^{-\mu_2\tau_3} e^{-\lambda_2\tau_3} \lambda_1 e^{-\lambda_1(t-\tau_1-\tau_2-\tau_3)} e^{-\mu_1(t-\tau_1-\tau_2-\tau_3)}$$

$$f_{3,4}(t) = \int_0^t d\tau_1 f_{2,3}(\tau_1) \lambda_2 e^{-\lambda_2(t-\tau_1)} e^{-\mu_2(t-\tau_1)} + \int_0^t d\tau_1 f_{3,3}(\tau_1) \int_0^{t-\tau_1} d\tau_2 \mu_3 e^{-\mu_3\tau_2} e^{-\lambda_3\tau_2} \lambda_2 e^{-\lambda_2(t-\tau_1-\tau_2)} e^{-\mu_2(t-\tau_1-\tau_2)}$$

$$f_{4,4}(t) = \int_0^t d\tau_1 f_{3,3}(\tau_1) \lambda_3 e^{-\lambda_3(t-\tau_1)} e^{-\mu_3(t-\tau_1)},$$

the LTs of which are given by

$$\hat{f}_{1,4}(s) = \hat{f}_{1,3}(s) \frac{\mu_1}{s+\mu_1+\lambda_1} \frac{\lambda_0}{s+\lambda_0} + \hat{f}_{2,3}(s) \frac{\mu_2}{s+\mu_2+\lambda_2} \frac{\mu_1}{s+\mu_1+\lambda_1} \frac{\lambda_0}{s+\lambda_0}$$

$$+ \hat{f}_{3,3}(s) \frac{\mu_3}{s+\mu_3+\lambda_3} \frac{\mu_2}{s+\mu_2+\lambda_2} \frac{\mu_1}{s+\mu_1+\lambda_1} \frac{\lambda_0}{s+\lambda_0}$$

$$\hat{f}_{2,4}(s) = \hat{f}_{1,3}(s) \frac{\lambda_1}{s+\mu_1+\lambda_1} + \hat{f}_{2,3}(s) \frac{\mu_2}{s+\mu_2+\lambda_2} \frac{\mu_1}{s+\mu_1+\lambda_1}$$

$$+ \hat{f}_{3,3}(s) \frac{\mu_3}{s+\mu_3+\lambda_3} \frac{\mu_2}{s+\mu_2+\lambda_2} \frac{\lambda_1}{s+\mu_1+\lambda_1}$$

$$\hat{f}_{3,4}(s) = \hat{f}_{2,3}(s)\frac{\lambda_2}{s+\mu_2+\lambda_2} + \hat{f}_{3,3}(s)\frac{\mu_3}{s+\mu_3+\lambda_3}\frac{\lambda_2}{s+\mu_2+\lambda_2}$$

$$\hat{f}_{4,4}(s) = \hat{f}_{3,3}(s)\frac{\lambda_3}{s+\mu_3+\lambda_3}.$$

It is straightforward to generalize the above and obtain the LT of $f_{k,m}(t)$, the probability density function for the time to complete the $m$th birth ($m = 2, 3,\ldots$) with the system size $k$ (= 1, 2, …, $m$) at the moment of the $m$th production. The LT of $f_{k,m}(t)$ satisfies the recursion relation:

$$\hat{f}_{k,m}(s) = \hat{f}_{k-1,m-1}(s)\frac{\lambda_{k-1}}{s+\mu_{k-1}+\lambda_{k-1}} + \sum_{j=0}^{m-1-k}\hat{f}_{k+j,m-1}(s)\left(\prod_{l=0}^{j}\frac{\mu_{k+j-l}}{s+\mu_{k+j-l}+\lambda_{k+j-l}}\right)\frac{\lambda_{k-1}}{s+\mu_{k-1}+\lambda_{k-1}} \quad \text{(A-1)}$$

with $\hat{f}_{1,1}(s) = \dfrac{\lambda_0}{s+\lambda_0}$, which can be proved by means of mathematical induction. Equation (A-1) indicates that, in order to have $k$ products at the moment of the $m$th birth, the number of products should be greater than or equal to $k-1$ and less than or equal to $m-1$ at the moment of the $(m-1)$th birth. Similarly, in order to have $k$ products at the moment of the $(m+1)$th birth, the number of products should be neither less than $k-1$ nor greater than $m$ at the moment of the $m$th birth. Specifically, when $m = 2$, Eq. (A-1) reduces to $\hat{f}_{1,2}(s) = \hat{f}_{1,1}(s)\dfrac{\mu_1}{s+\mu_1+\lambda_1}\dfrac{\lambda_0}{s+\lambda_0}$ and $\hat{f}_{2,2}(s) = \hat{f}_{1,1}(s)\dfrac{\lambda_1}{s+\mu_1+\lambda_1}$. Furthermore, assuming that Eq. (A-1) is valid for $m = n$, we can see its validity for $m = n + 1$ as well. Unless any death occurs, Eq. (A-1) reduces to $\hat{f}_{m,m}(s) = \hat{f}_{m-1,m-1}(s)\dfrac{\lambda_{m-1}}{s+\lambda_{m-1}}$ or $f_{m,m}(t) = \int_0^t d\tau\, f_{m-1,m-1}(\tau)\lambda_{m-1}e^{-\lambda_{m-1}(t-\tau)}$ because $\mu_j = 0$ for all $j = 1, 2, \ldots$, and $f_{k,m}(t) = 0$ for $k = 1, 2, \cdots, m-1$. The inverse Laplace transform (ILT) of $\hat{f}(s)$ is given as the Bromwich integral $f(t) = \dfrac{1}{2\pi i}\int_{\alpha-i\infty}^{\alpha+i\infty} ds\, \hat{f}(s)e^{st}$, where $\alpha$ is a positive constant large enough that all singularities of $\hat{f}(s)$ lie on the left of a vertical line segment from $s = \alpha - i\infty$ to

$s = \alpha + i\infty$. The Bromwich integral can be written in the form $f(t) = \sum_{n=1}^{N} \text{Res}_{s=s_n} \left[ e^{st} \hat{f}(s) \right]$, where $s_n$ (for $n = 1, 2, \ldots, N$) denotes the singularities of $\hat{f}(s)$ and $\text{Res}_{z=z_0} f(z)$ represents the residue of $f(z)$ at the singular point $z = z_0$. Using the Bromwich integral, we take the ILT of $\prod_{l=0}^{j}(s+a_l)^{-1}$ before determining the analytic expression of ILT of Eq. (A-1). If $(s+a_i)^{-1}$ is an $m_i$-fold multiplication, we can generally write $\prod_{l=1}^{j}(s+a_l)^{-1} = \left( \prod_{i=1}^{q}(s+a_i)^{-m_i} \right) \prod_{r=\sum_{i=1}^{q} m_i +1}^{j} (s+a_r)^{-1}$ and obtain

$$L^{-1}\left( \prod_{l=0}^{j}(s+a_l)^{-1} \right) = \sum_{i=1}^{q} \frac{1}{(m_i-1)!} \frac{\partial^{m_i-1}}{\partial s^{m_i-1}} \left\{ \left( \prod_{\substack{u=1 \\ u \neq i}}^{q}(s+a_u)^{-m_u} \right) \left( \prod_{r=\sum_{i=1}^{q} m_i +1}^{j+1}(s+a_r)^{-1} \right) e^{st} \right\} \Bigg|_{s=-a_i}$$

$$+ \sum_{r=\sum_{i=1}^{q} m_i +1}^{j+1} \left\{ \left( \prod_{i=0}^{q}(a_i-a_r)^{-m_i} \right) \left( \prod_{\substack{v=\sum_{i=1}^{q} m_i +1 \\ v \neq r}}^{j+1}(a_v-a_r)^{-1} \right) e^{-a_r t} \right\}$$

where $L^{-1}$ denotes the ILT. In this formula, the differentiation of the first term on the right-hand side can be rewritten in terms of Leibniz's rule. Therefore, the ILT of Eq. (A-1) takes the form

$$f_{k,m}(t) = \int_0^t d\tau\, f_{k-1,m-1}(\tau) \lambda_{k-1} e^{-(\mu_{k-1}+\lambda_{k-1})(t-\tau)}$$
$$+ \sum_{j=0}^{m-1-k} \left[ \left( \prod_{l=0}^{j} \mu_{k+j-l} \right) \int_0^t d\tau_1\, f_{k+j,m-1}(\tau_1) \int_0^{t-\tau_1} d\tau_2\, g_{k,j}(\tau_2) \lambda_{k-1} e^{-(\mu_{k-1}+\lambda_{k-1})(t-\tau_1-\tau_2)} \right] \quad \text{(A-2)}$$

with

$$g_{k,j}(\tau_2) \equiv L^{-1}\left(\prod_{l=0}^{j}\left(s+\mu_{k+j-l}+\lambda_{k+j-l}\right)^{-1}\right)$$

$$= \sum_{i=1}^{q}\sum_{z_1}\sum_{z_2}\cdots\sum_{z_{i-1}}\sum_{z_{i+1}}\cdots\sum_{z_q}\sum_{z_{q+1}}\cdots\sum_{z_{q+j-\sum_{i=1}^{q}m_i+2}} \left\{ \frac{\prod_{\substack{u=1\\u\neq i}}^{q}(-1)^{z_u}\frac{(m_u+z_u-1)!}{(m_u-1)!}(c_{k+j,u}-c_{k+j,i})^{-m_u-z_u}}{z_1!z_2!\cdots z_{i-1}!z_{i+1}!\cdots z_q!z_{q+1}!\cdots z_{q+j-\sum_{i=1}^{q}m_i+1}!} \right.$$

$$\left. \times \left(\prod_{r=q+1}^{q+j-\sum_{i=1}^{q}m_i+1}(-1)^{z_r}(z_r!)(c_{k+j,r}-c_{k+j,i})^{-1-z_r}\right) \tau_2^{z_{q+j-\sum_{i=1}^{q}m_i+2}} e^{-c_{k+j,i}\tau_2} \right\}$$

$$+ \sum_{r=q+1}^{q+1+j-\sum_{i=1}^{q}m_i} \left\{\left(\prod_{i=1}^{q}(c_{k+j,i}-c_{k+j,r})^{-m_i}\right)\left(\prod_{\substack{v=q+1\\v\neq r}}^{q+1+j-\sum_{i=1}^{q}m_i}(c_{k+j,v}-c_{k+j,r})^{-1}\right)e^{-c_{k+j,r}\tau_2}\right\},$$

where $z_w$'s are integers $w=1,2,\cdots,q+j-\sum_{i=1}^{q}m_i+1$, ranging from 0 to $m_i-1$ (i.e., $0\leq z_w\leq m_i-1$)

and satisfying $\sum_{\substack{w=1\\w\neq i}}^{q+j-\sum_{i=1}^{q}m_i+1} z_w = m_i-1$ $(i=1,2,\cdots,q)$, and $c_{k+j,u}$ is defined to be one of

$\mu_{k+j-l}+\lambda_{k+j-l}$ $(l=0,1,2,\cdots,j)$. For $a_u$'s and $c_y$'s are ordered according to multiplicity from the smallest to the largest. For example, we order $a_1=1$, $a_2=3$, $a_3=1$, $a_4=4$, $a_5=2$, $a_6=3$, and $a_7=5$ as $c_1=1=a_1=a_3$, $c_2=3=a_2=a_6$, $c_3=2=a_5$, $c_4=4=a_4$, and $c_5=5=a_7$. If birth and death rates are all different from each other, namely, $\lambda_i\neq\lambda_j$ and $\mu_i\neq\mu_j$ for any $i$ and $j$ ($\neq i$). Further, since $q$ is zero, it follows that $g_{k,j}(\tau_2) = \sum_{r=1}^{j+1}\left(\prod_{\substack{v=1\\v\neq r}}^{j+1}(c_{k+j,v}-c_{k+j,r})^{-1}\right)e^{-c_{k+j,r}\tau_2}$.

Equipped with these, we are now ready to derive $P_n(t)$ from Eq. (A-1). We categorize the case of the system being empty (i.e., the system size being zero) at time $t$ into two kinds: One is that the first birth does not occur until time $t$, and the other is that all born products are destroyed, resulting in none to exist at time $t$. The probability that the first birth would not occur until time $t$ reads $e^{-\lambda_0 t}$, the LT of which is $(s+\lambda_0)^{-1}$. Let us consider the case of the empty system in the presence of birth occurring at least once. Suppose that $k$ products exist when the $m$th birth occurs at time $\tau_1$ and the $k$ products are destroyed during the interval $\tau_2-\tau_1$. To have none at time $t$, we should have no birth until $t$. This case is described by

$$\sum_{m=1}^{\infty}\sum_{k=1}^{m} \hat{f}_{k,m}(s)\left(\prod_{l=0}^{k-1}\frac{\mu_{k-l}}{s+\mu_{k-l}+\lambda_{k-l}}\right)\frac{1}{s+\lambda_0},$$

which is the LT of the corresponding time expression at $s$. Then, we have $\hat{P}_0(s) = \frac{1}{s+\lambda_0} + \sum_{m=1}^{\infty}\sum_{k=1}^{m} \hat{f}_{k,m}(s)\left(\prod_{l=0}^{k-1}\frac{\mu_{k-l}}{s+\mu_{k-l}+\lambda_{k-l}}\right)\frac{1}{s+\lambda_0}$, where the first term $\frac{1}{s+\lambda_0}$ simply equals $\frac{1}{\lambda_0}\hat{f}_{1,1}(s)$. It is also straightforward to show that the summation term

$$\sum_{m=1}^{\infty}\sum_{k=1}^{m} \hat{f}_{k,m}(s)\left(\prod_{l=0}^{k-1}\frac{\mu_{k-l}}{s+\mu_{k-l}+\lambda_{k-l}}\right)\frac{1}{s+\lambda_0} \text{ reduces to } \frac{1}{\lambda_0}\sum_{m=2}^{\infty}\hat{f}_{1,m}(s) \text{ since putting } k=1 \text{ in Eq. (A-1)}$$

results in $\hat{f}_{1,m}(s) = \sum_{j=0}^{m-2}\hat{f}_{j+1,m-1}(s)\left(\prod_{l=0}^{j}\frac{\mu_{j+1-l}}{s+\mu_{j+1-l}+\lambda_{j+1-l}}\right)\frac{\lambda_0}{s+\lambda_0}$. As consequence $\hat{P}_0(s)$ obtains a considerably simpler expression: $\hat{P}_0(s) = \frac{1}{\lambda_0}\sum_{m=1}^{\infty}\hat{f}_{1,m}(s)$. As for $P_1(t)$, there should be at least one birth involved until time $t$. If there are $k$ products when the $m$th birth occurs at time $\tau$, we can have one product at time $t$ if $k-1$ products among the $k$ products destruct with no birth during $t-\tau$. Considering this argument, we obtain $\hat{P}_1(s) = \sum_{m=1}^{\infty}\sum_{k=1}^{m}\hat{f}_{k,m}(s)\left(\prod_{l=0}^{k-2}\frac{\mu_{k-l}}{s+\mu_{k-l}+\lambda_{k-l}}\right)\frac{1}{s+\mu_1+\lambda_1}$.

Similarly, the LT of the probability that the system size is two at time $t$ is given by

$\hat{P}_2(s) = \sum_{m=2}^{\infty}\sum_{k=2}^{m}\hat{f}_{k,m}(s)\left(\prod_{l=0}^{k-3}\frac{\mu_{k-l}}{s+\mu_{k-l}+\lambda_{k-l}}\right)\frac{1}{s+\mu_2+\lambda_2}$. Note that $\hat{P}_1(s)$ and $\hat{P}_2(s)$ also reduce to simpler expressions: $\hat{P}_1(s) = \frac{1}{\lambda_1}\sum_{m=2}^{\infty}\hat{f}_{2,m}(s)$ and $\hat{P}_2(s) = \frac{1}{\lambda_2}\sum_{m=3}^{\infty}\hat{f}_{3,m}(s)$. Generalizing the result of $\hat{P}_n(s)$ for non-negative integer $n$ (= 0, 1, 2, ...), we complete the proof of Eq. (2). Without destruction ($\mu_n = 0$), Eq. (2) obtains the form $P_n(t) = \frac{1}{\lambda_n}f_{n+1,n+1}(t)$, which is consistent with the case of only birth. Then the sum of $\hat{P}_n(s)$ over $n$ is rearranged as follows:

$$\sum_{n=0}^{\infty} \hat{P}_n(s) = \frac{1}{s} - \frac{\hat{f}_{1,1}(s)}{s} + \left( \frac{\hat{f}_{1,1}(s)}{s} - \frac{\sum_{i=1}^{2} \hat{f}_{i,2}(s)}{s} \right) + \left( \frac{\sum_{i=1}^{2} \hat{f}_{i,2}(s)}{s} - \frac{\sum_{i=1}^{3} \hat{f}_{i,3}(s)}{s} \right) + \cdots,$$

which completes the proof of the normalization condition: $\sum_{n=0}^{\infty} \hat{P}_n(s) = \frac{1}{s}$.

We now confirm that $\hat{P}_n(s) = \frac{1}{\lambda_n} \sum_{m=n+1}^{\infty} \hat{f}_{n+1,m}(s)$ satisfies Eq. (1). Examining the relation between $\hat{P}_0(s)$ and $\hat{P}_1(s)$, we find $(s + \lambda_0)\hat{P}_0(s) - 1 = \gamma_1 \hat{P}_1(s)$ or $s\hat{P}_0(s) - P_0(0) = -\lambda_0 \hat{P}_0(s) + \mu_1 \hat{P}_1(s)$ due to the initial condition $P_n(0) = \delta_{n0}$. Similarly, it is straightforward to show that $\hat{P}_{n-1}(s)$, $\hat{P}_n(s)$, and $\hat{P}_{n+1}(s)$ are related via $s\hat{P}_n(s) = -(\lambda_n + \mu_n)\hat{P}_n(s) + \lambda_{n-1}\hat{P}_{n-1}(s) + \mu_{n+1}\hat{P}_{n+1}(s)$ for $n \geq 1$. The ILT then leads to Eq. (1), which manifests that Eq. (2) is the solution of Eq. (1).

2. *Finite system capacity and an unlimited number of death channels*

Next, we focus on the birth-death system in which the number of death channels is still arbitrarily large, but a limit $N$ is placed on the number allowed in the system at any time. The procedure is the same as that for the infinite system capacity, except that the birth rate $\lambda_n$ vanishes whenever $n \geq N$. It then follows from Eq. (A-1) that the LT of the probability rate function $f_{k,m}(t)$ for the time to complete the $m$th birth in the system of size $k$ is given by

$$\hat{f}_{k,m}(s) = \begin{cases} \hat{f}_{k-1,m-1}(s) \dfrac{\lambda_{k-1}}{s+\mu_{k-1}+\lambda_{k-1}} \\ \quad + \displaystyle\sum_{j=0}^{m-1-k} \hat{f}_{k+j,m-1}(s) \left( \prod_{q=0}^{j} \dfrac{\mu_{k+j-q}}{s+\mu_{k+j-q}+\lambda_{k+j-q}} \right) \dfrac{\lambda_{k-1}}{s+\mu_{k-1}+\lambda_{k-1}} \quad (m \le N) \\ \hat{f}_{k-1,m-1}(s) \dfrac{\lambda_{k-1}}{s+\mu_{k-1}+\lambda_{k-1}} \\ \quad + \displaystyle\sum_{j=0}^{N-k} \hat{f}_{k+j,m-1}(s) \dfrac{\mu_k}{s+\mu_k} \left( \prod_{q=1}^{j} \dfrac{\mu_{k+j-q}}{s+\mu_{k+j-q}+\lambda_{k+j-q}} \right) \dfrac{\lambda_{k-1}}{s+\mu_{k-1}+\lambda_{k-1}} \quad (m > N) \end{cases}$$

(A-3)

The first equality of Eq. (A-3) is identical to Eq. (A-1). The final term in the second equality follows because the system capacity is $N$. Despite the differences between Eqs. (A-1) and (A-3), the relationship does not change between the probability $P_n(t)$ and $f_{k,m}(t)$. In other words, Eq. (3) is still suitable for the birth-death system with capacity limitations. Moreover, Eq. (A-3) also leads to the fulfilment of the normalization condition $\sum_{n=0}^{N} P_n(t) = 1$.

3. *Infinite system capacity and a finite number of death channels*

Here we consider a birth-death system with death channels of finite number $c$ and infinite system capacity. It is helpful to distinguish the rate of destruction depending on the number $n$ of products in the system. If there are $c$ or more products, the destruction rate is $\mu_c$ because all $c$ channels must be busy. When products less than $c$ exist in the system ($n < c$), only $n$ of the $c$ channels should be in service, yielding the destruction rate $\mu_n$. In consequence, the LT of $f_{k,m}(t)$ is written as follows:

$$\hat{f}_{k,m}(s) = \begin{cases} \hat{f}_{k-1,m-1}(s)\dfrac{\lambda_{k-1}}{s+\mu_{k-1}+\lambda_{k-1}} \\ \quad + \displaystyle\sum_{j=0}^{m-1-k} \hat{f}_{k+j,m-1}(s)\left(\prod_{l=0}^{j}\dfrac{\mu_{k+j-l}}{s+\mu_{k+j-l}+\lambda_{k+j-l}}\right)\dfrac{\lambda_{k-1}}{s+\mu_{k-1}+\lambda_{k-1}} & (m \leq c+1) \\[2ex] \hat{f}_{k-1,m-1}(s)\dfrac{\lambda_{k-1}}{s+\mu_{k-1}+\lambda_{k-1}} \\ \quad + \displaystyle\sum_{j=0}^{c-k} \hat{f}_{k+j,m-1}(s)\left(\prod_{v=0}^{j}\dfrac{\mu_{k+j-v}}{s+\mu_{k+j-v}+\lambda_{k+j-v}}\right)\dfrac{\lambda_{k-1}}{s+\mu_{k-1}+\lambda_{k-1}} \\ \quad + \displaystyle\sum_{j=c-k+1}^{c+m-1-k} \hat{f}_{k+j,m-1}(s)\left(\prod_{q=1}^{k+j-c}\dfrac{\mu_c}{s+\mu_c+\lambda_{c+q}}\right)\left(\prod_{r=0}^{c-k}\dfrac{\mu_{c-r}}{s+\mu_{c-r}+\lambda_{c-r}}\right)\dfrac{\lambda_{k-1}}{s+\mu_{k-1}+\lambda_{k-1}} \\ & (m > c+1 \text{ and } k \leq c) \\[2ex] \hat{f}_{k-1,m-1}(s)\dfrac{\lambda_{k-1}}{s+\mu_{k-1}+\lambda_{k-1}} \\ \quad + \displaystyle\sum_{j=0}^{c+m-1-k} \hat{f}_{k+j,m-1}(s)\left(\prod_{q=0}^{j}\dfrac{\mu_c}{s+\mu_c+\lambda_{k+q}}\right)\dfrac{\lambda_{k-1}}{s+\mu_{k-1}+\lambda_{k-1}} & (m < c+1 \text{ and } k > c) \end{cases}$$

. (A-4)

While the first equality of Eq. (A-4) is the same as Eq. (A-1), the second and third equalities result from the destruction rate given by $\mu_c$ for the population larger than the finite number $c$ of death channels.

### 4. *Finite system capacity and a finite number of death channels*

Finally, we examine a birth-death system in which both the system capacity and the number of death channels are finite, say, $N$ and $c$, respectively. This birth-death system can be classified into two groups: $N \leq c$ and $N > c$. The case $N \leq c$ is essentially the same as the case of finite system capacity $N$ and the infinite number of death channels. We, therefore, consider only the case $N > c$ and use the same procedure as the system with infinite system capacity and a finite number of channels, except that the production rate vanishes ($\lambda_n = 0$) for $n \geq N$. It then follows from Eq. (A-4) that the LT of $f_{k,m}(t)$ is given by

$$\hat{f}_{k,m}(s) = \begin{cases} \hat{f}_{k-1,m-1}(s)\dfrac{\lambda_{k-1}}{s+\mu_{k-1}+\lambda_{k-1}} \\ \quad + \sum_{j=0}^{m-1-k} \hat{f}_{k+j,m-1}(s)\left(\prod_{l=0}^{j}\dfrac{\mu_{k+j-l}}{s+\mu_{k+j-l}+\lambda_{k+j-l}}\right)\dfrac{\lambda_{k-1}}{s+\mu_{k-1}+\lambda_{k-1}} \quad (m \le c+1) \\[2mm] \hat{f}_{k-1,m-1}(s)\dfrac{\lambda_{k-1}}{s+\mu_{k-1}+\lambda_{k-1}} \\ \quad + \sum_{j=0}^{c-k} \hat{f}_{k+j,m-1}(s)\left(\prod_{v=0}^{j}\dfrac{\mu_{k+j-v}}{s+\mu_{k+j-v}+\lambda_{k+j-v}}\right)\dfrac{\lambda_{k-1}}{s+\mu_{k-1}+\lambda_{k-1}} \\ \quad + \sum_{j=c-k+1}^{\min[N,c+m-1]-k} \hat{f}_{k+j,m-1}(s)\left(\prod_{q=1}^{k+j-c}\dfrac{\mu_c}{s+\mu_c+\lambda_{c+q}}\right)\left(\prod_{r=0}^{c-k}\dfrac{\mu_{c-r}}{s+\mu_{c-r}+\lambda_{c-r}}\right)\dfrac{\lambda_{k-1}}{s+\mu_{k-1}+\lambda_{k-1}} \\ \hspace{6cm} (m > c+1 \text{ and } k \le c) \\[2mm] \hat{f}_{k-1,m-1}(s)\dfrac{\lambda_{k-1}}{s+\mu_{k-1}+\lambda_{k-1}} \\ \quad + \sum_{j=0}^{\min[N,c+m-1]-k} \hat{f}_{k+j,m-1}(s)\left(\prod_{q=0}^{j}\dfrac{\mu_c}{s+\mu_c+\lambda_{k+q}}\right)\dfrac{\lambda_{k-1}}{s+\mu_{k-1}+\lambda_{k-1}} \quad (m > c+1 \text{ and } k > c) \end{cases}$$

, (A-5)

where min[$a$, $b$] in the upper bound of the last summation in the second and third equalities stands for the minimum value of $a$ and $b$. Note that Eq. (A-5) is a mixture of Eqs. (A-3) and (A-4).

We also point out that the first terms in Eqs. (A-3), (A-4), and (A-5) are identical to that in Eq. (A-1). Despite the differences between Eq. (A-1) and Eqs. (A-3), (A-4), and (A-5), the relationship between the probability $P_n(t)$ and $f_{k,m}(t)$ remains unchanged. In other words, Eq. (3) is still suitable for the birth-death system regardless of the system capacity and the number of death channels. Further, Eqs. (A-3), (A-4), and (A-5) also lead to the probability $P_n(t)$ that fulfils the normalization conditions $\sum_{n=0}^{N} P_n(t) = 1$, $\sum_{n=0}^{\infty} P_n(t) = 1$, and $\sum_{n=0}^{\max[N,c]} P_n(t) = 1$, respectively.

**Derivation of Eqs. (6) and (7)**

It is straightforward to derive Eqs. (6) and (7). Multiplying both sides of Eq. (1) by $n$ and summing over $n$ from zero to infinity, we obtain

$$\frac{\partial \langle n(t) \rangle}{\partial t} \equiv \frac{\partial}{\partial t} \sum_{n=0}^{\infty} n P_n(t) = -\sum_{n=0}^{\infty} n(\lambda_n + \mu_n) P_n(t) + \sum_{n=0}^{\infty} n\lambda_{n-1} P_{n-1}(t) + \sum_{n=0}^{\infty} n\mu_{n+1} P_{n+1}(t)$$

$$= -[\langle n\lambda_n(t) \rangle + \langle n\mu_n(t) \rangle] + \sum_{m=0}^{\infty} (m+1)\lambda_m P_m(t) + \sum_{m=0}^{\infty} (m-1)\mu_m P_m(t) \quad . \quad \text{(A-6)}$$

$$= \sum_{n=0}^{\infty} (\lambda_n - \mu_n) P_n(t) = \langle \lambda_n(t) \rangle - \langle \mu_n(t) \rangle$$

Integration of Eq. (A-6) over $t$ completes the proof of Eq. (6). To obtain Eq. (7), we should first determine the second moment of the product number. Multiplying Eq. (1) by $n^2$ and summing over $n$ yield

$$\frac{\partial \langle n^2(t) \rangle}{\partial t} = -\left[\langle n^2\lambda_n(t) \rangle + \langle n^2\mu_n(t) \rangle\right] + \sum_{n=0}^{\infty} n^2\lambda_{n-1} P_{n-1}(t) + \sum_{n=0}^{\infty} n^2\mu_{n+1} P_{n+1}(t)$$

$$= -\left[\langle n^2\lambda_n(t) \rangle + \langle n^2\mu_n(t) \rangle\right] + \sum_{m=0}^{\infty} (m+1)^2 \lambda_m P_m(t) + \sum_{m=0}^{\infty} (m-1)^2 \mu_m P_m(t)$$

$$= -\left[\langle n^2\lambda_n(t) \rangle + \langle n^2\mu_n(t) \rangle\right] + \sum_{m=0}^{\infty} \left\{(m^2+2m+1)\lambda_m P_m(t) + (m^2-2m+1)\mu_m P_m(t)\right\}. \quad \text{(A-7)}$$

$$= 2[\langle n\lambda_n(t) \rangle - \langle n\mu_n(t) \rangle] + \langle \lambda_n(t) \rangle + \langle \mu_n(t) \rangle$$

$$= 2[\langle n\lambda_n(t) \rangle - \langle n\mu_n(t) \rangle] + \frac{\partial \langle n(t) \rangle}{\partial t} + 2\langle \mu_n(t) \rangle$$

Integrating both sides of Eq. (A-7), we obtain

$$\langle n^2(t) \rangle = \langle n(t) \rangle + 2\int_0^t d\tau \left[\langle n\lambda_n(\tau) \rangle - \langle n\mu_n(\tau) \rangle + \langle \mu_n(\tau) \rangle\right]$$

$$= \langle n(t) \rangle + 2\int_0^t d\tau \left\{\text{Cov}(n, \lambda_n(\tau)) + \langle n(\tau) \rangle \langle \lambda_n(\tau) \rangle - \text{Cov}[n, \mu_n(\tau)] - \langle n(\tau) \rangle \langle \mu_n(\tau) \rangle + \langle \mu_n(\tau) \rangle\right\}$$

$$= \langle n(t) \rangle + 2\int_0^t d\tau \left\{\text{Cov}[n, \lambda_n(\tau) - \mu_n(\tau)] + \langle n(\tau) \rangle \frac{d\langle n(\tau) \rangle}{d\tau} + \langle \mu_n(\tau) \rangle\right\}$$

$$= \langle n(t) \rangle + \langle n(t) \rangle^2 + 2\int_0^t d\tau \left\{\text{Cov}[n, \lambda_n(\tau) - \mu_n(\tau)] + \langle \mu_n(\tau) \rangle\right\}$$

,

(A-8)

where $\mathrm{Cov}[n, X(\tau)] \equiv \langle nX(\tau) \rangle - \langle n(\tau) \rangle \langle X(\tau) \rangle$ is the covariance of *n* and *X*. The third equality of Eq. (A-8) follows from Eq. (A-6). Thus, the variance of the system size obtains the form:

$$\sigma_n^2(t) = \langle n(t) \rangle + 2\int_0^t d\tau \left\{ \mathrm{Cov}[n, \lambda_n(\tau) - \mu_n(\tau)] + \langle \mu_n(\tau) \rangle \right\} . \tag{A-9}$$

Here the covariance of *n* and $\lambda_n(t) - \mu_n(t)$ is given by the correlation coefficient $\rho(t)$ of the two variables multiplied by the product of their standard deviations $\sigma_n(t)\sigma_{\lambda_n - \mu_n}(t)$: $\mathrm{Cov}[n, \lambda_n(t) - \mu_n(t)] = \rho(t)\sigma_n(t)\sigma_{\lambda_n - \mu_n}(t)$, where $\rho(t)$ ranges from −1 to 1. The positive values of $\rho(t)(>0)$ imply that the birth rate minus the death rate $\lambda_n(t) - \mu_n(t)$ in the presence of *n* products at time *t* increases with the population *n*. The negative values of $\rho(t)(<0)$ imply that $\lambda_n(t) - \mu_n(t)$ decreases as *n* grows. The null correlation coefficient $[\rho(t) = 0]$ reflects that $\lambda_n(t) - \mu_n(t)$ is independent of *n*. Substituting $\mathrm{Cov}[n, \lambda_n(t) - \mu_n(t)] = \rho(t)\sigma_n(t)\sigma_{\lambda_n - \mu_n}(t)$ into Eq. (A-9), we finally obtain

$$\sigma_n^2(t) = \langle n(t) \rangle + 2\int_0^t d\tau \left[ \rho(\tau)\sigma_n(\tau)\sigma_{\lambda_n - \mu_n}(\tau) + \langle \mu_n(\tau) \rangle \right],$$

which completes the proof of Eq. (7).